\documentclass[12pt]{iopart}
\usepackage{graphicx}
\begin{document}
\title[]{Modification of the trapped field in bulk high-temperature superconductors as a result of the drilling of a pattern of artificial columnar holes}

\author{Gregory P. Lousberg$^{1,4}$, J-F Fagnard$^{2}$, M Ausloos$^{3}$, Ph~Vanderbemden$^{1}$ and B Vanderheyden$^{1}$}
\address{$^{1}$ SUPRATECS Research Group, Dept. of Electrical Engineering and Computer Science (B28), University of Li\`ege, Belgium}
\address{$^{2}$ CISS Department, Royal Military Academy, Belgium}
\address{$^{3}$ SUPRATECS (B5a), University of Li\`ege, Belgium}
\address{$^{4}$ FRS-FNRS fellowship}

\ead{gregory.lousberg@ulg.ac.be}

\begin{abstract}
The trapped magnetic field is examined in bulk high-temperature superconductors that are artificially drilled along their $c$-axis. The influence of the hole pattern on the magnetization is studied and compared by means of numerical models and Hall probe mapping techniques. To this aim, we consider two bulk YBCO samples with a rectangular cross-section that are drilled each by six holes arranged either on a rectangular lattice (sample I) or on a centered rectangular lattice (sample II). For the numerical analysis, three different models are considered for calculating the trapped flux: (i), a two-dimensional (2D) Bean model neglecting demagnetizing effects and flux creep, (ii), a 2D finite-element model neglecting demagnetizing effects but incorporating magnetic relaxation in the form of an $E-J$ power law, and, (iii), a 3D finite element analysis that takes into account both the finite height of the sample and flux creep effects. For the experimental analysis, the trapped magnetic flux density is measured above the sample surface by Hall probe mapping performed before and after the drilling process. The maximum trapped flux density in the drilled samples is found to be smaller than that in the plain samples. The smallest magnetization drop is found for sample II, with the centered rectangular lattice. This result is confirmed by the numerical models. In each sample, the relative drops that are calculated independently with the three different models are in good agreement. As observed experimentally, the magnetization drop calculated in the sample II is the smallest one and its relative value is comparable to the measured one. By contrast, the measured magnetization drop in sample (1) is much larger than that predicted by the simulations, most likely because of a change of the microstructure during the drilling process.

\end{abstract}

\section{Introduction}

Single crystals of YBa$_2$Cu$_3$O$_{7-\delta}$ containing a series of holes that are artificially drilled along a given crystallographic direction have been recently introduced in order to improve the performances of YBCO trapped field magnets~\cite{1,2,3,4,Supratecs}. For instance, it was demonstrated that the maximum trapped flux density in YBCO magnets may be enhanced in the presence of holes that favor the oxygen annealing and thus yield larger critical current densities~\cite{FirstDrilled,10,11,14}. Moreover, when subjected to variable magnetic fields, the drilled trapped field magnets exhibit a better thermal stability than non-drilled magnets, since they offer a larger surface for exchanging heat with the environment. As a result, losses associated with the thermal motion of vortices are reduced~\cite{12}. Last, by impregnating the holes with resin, the drilled magnets can be reinforced against the strong Lorentz forces induced in the bulk of the material~\cite{9}.

As regards the magnetic properties, the holes of a drilled sample influence the current stream lines --- which cannot cross them --- and thus affect the magnetic field distribution.  In previous works, we modelled the current stream lines in the presence of holes and calculated the trapped field in drilled samples of finite or infinite heights~\cite{Lousberg, Lousberg2}. It was shown that, for a given critical current density, the maximum of the trapped magnetic field is always smaller in a drilled sample than in a non-drilled one. Moreover, it was demonstrated that for a given sample geometry, there is an optimal position of the holes which minimizes the drop of trapped flux. The holes must be arranged in such a manner that their centre lies on the discontinuity lines\footnote{Discontinuity lines are lines near a given hole and across which the current density changes abruptly its direction.} of the neighboring holes.

In the literature, although it is has been already demonstrated that drilling holes in a sample decreases its trapped flux~\cite{15,16}, no experimental study has been reported so far about the influence of the hole pattern on the trapping properties of drilled samples. The purpose of this paper is to show experimentally that, considering a given number of identical holes, their geometrical arrangement has an influence on the maximum trapped flux density. These measurements are confronted to the theoretical predictions made in Refs.~\cite{Lousberg, Lousberg2}.

The paper is organized as follows. Section~\ref{s:sample} describes the preparation and the properties of the samples under study. In Section~\ref{s:models}, we describe the numerical models of increasing complexity (taking progressively into account flux creep and geometrical effects) to predict the trapped flux in the samples. Section~\ref{s:measure} reports on the measurements of the trapped flux before and after the drilling of the holes, and Section~\ref{s:discussion} discuss them in the light of results obtained with the numerical models. Section~\ref{s:concl} presents the conclusions.

\section{Sample preparation}
\label{s:sample}

In most cases considered so far~\cite{FirstDrilled,10,11,14}, drilled samples were obtained from Y211-preform where the holes had been arranged \emph{before} the synthesis of the Y123 phase. This approach makes the comparison of drilled samples with different hole arrangements difficult, since the critical current densities may differ appreciably from one sample to another even though they are processed similarly. We thus followed a different approach and considered bulk YBCO samples that were drilled \emph{after} their synthesis. This approach allowed us to compare the trapped magnetic flux in every sample before and after drilling the holes. Provided the microstructure of the sample has not been affected by the drilling process, one can make the reasonable assumption that the critical current density is unchanged and consider the sole effect of the modification of the current stream lines on the magnetic properties of the sample.

\begin{figure}[t]
\center
\includegraphics[width=14cm]{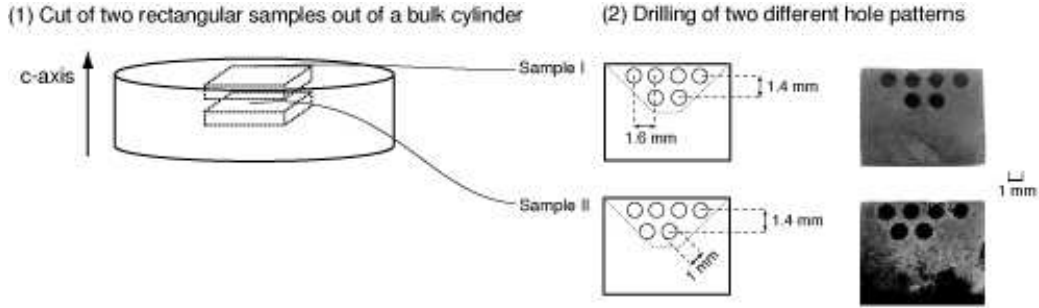}\caption{(1)- Extraction of two rectangular YBCO samples out of a cylindrical one, with the help of a wire saw. (2)- Sketch of the hole patterns drilled in the samples and pictures of the samples after the drilling process. }\label{FigureSketch}
\end{figure}

The plain samples were extracted from a top-seeded melt-grown YBCO bulk cylinder that was synthesized at CRISMAT (Caen, France). The cylinder has a diameter of $22~\mathrm{mm}$ and a height of $16~\mathrm{mm}$. With the help of a wire saw, two rectangular samples were cut out as sketched in Figure~\ref{FigureSketch}. The resulting samples have a length of $8.6~\mathrm{mm}$, a width of $6.6~\mathrm{mm}$, and a height of $4~\mathrm{mm}$.

In each sample, six holes with diameter $1~\mathrm{mm}$ were drilled along the $c$-axis. To prevent the superconducting ceramics from breaking during the drilling process, an ultrasonic drilling machine was used. In order to avoid effects associated with the finite side lengths of the samples, the six holes were made in the trapezoidal sector delimited by the gray dashed lines shown on Figure~\ref{FigureSketch}. For each sample, this sector lies in the rectangular cross section and is delimited by one long edge of the cross section, a segment of its longest median, and two lines which start at the corners with a 45 degree angle. This choice is motivated by the critical state configuration found in a non drilled sample subjected to a magnetic field applied along the $c$-axis. In this case, the two boundaries at 45 degrees are places where the current stream lines change abruptly their direction, whereas they flow parallel to the long edge within the trapezoidal sector~\cite{rectangle}. Thus, the holes are drilled in a sector where the current lines are initially parallel to one another, a situation which is close to that considered in the previous theoretical models~\cite{Lousberg}.

For each sample, two series of holes (containing 4 and 2 holes, respectively) are made along parallel lines. The hole separation along a line is $1.6~\mathrm{mm}$ and the lines are separated by $1.4~\mathrm{mm}$. The two holes on the second line can be arranged in two different ways: (I) either they are aligned with the holes of the first line and form the beginning of a {\it rectangular lattice}, or (II), the holes are shifted with respect to those of the first line by half a separation ($0.8~\mathrm{mm}$)  and start forming a {\it centered rectangular lattice}. In this latter case, the centres of the holes of the second line are in fact placed on the discontinuity lines associated with the holes of the first line~\cite{Lousberg}. A picture of the two samples obtained after the drilling process is shown in Figure~\ref{FigureSketch}.

\section{Numerical models}
\label{s:models}

The trapped magnetic flux density is calculated in samples I and II by means of three different numerical models of increasing complexity:
\begin{enumerate}

\item[(1)]{\bf A critical state model --- or Bean model ---}\\
In the first model, which is a generalisation of the Bean model for arbitrary cross sections~\cite{Bean, CampbellEvetts}, the samples are assumed to have an infinite height and a uniform critical current density, $J_c$. The gradient of magnetic flux density satisfies the relation
\begin{equation}
\frac{d\mathbf{B}}{d\ell}(P)=\left\{\begin{array}{c} \pm\mu_0 \mathbf{J_c} \quad \mbox{in penetrated regions},\\ 0 \quad \mbox{in virgin regions}, \end{array}\right.\label{Beaneq}
\end{equation}
where $P$ is a point at a given location, $\ell$ is the total length crossed by the external flux to reach  $P$, and $\mu_0$ is the magnetic permeability. This relation can be integrated numerically under the constraint that the penetration length $\ell$ be minimal at each point $P$ (or in other words, the shortest penetration path be selected for each point $P$)~\cite{Lousberg}. The magnetization can then be deduced from the resulting field distribution.
\item[(2)]{\bf 2D finite element model}\\
The second model also assumes that the samples have an infinite height but includes flux creep effects via the constitutive $E-J$ law 
\begin{equation}
\mathbf{E}=E_c\left(\frac{J}{J_c}\right)^n\:\frac{\mathbf{J}}{J},
\end{equation}
where $E_c$ is the critical electric field and $n$ is the critical exponent. The field distribution can then be solved by integrating the Maxwell equations. The model is solved numerically by a finite-element method (FEM) implemented in the {\it open source} solver GetDP~\cite{GetDP}. The detailed procedure has been described in Ref.~\cite{Lousberg2} and, for large values of $n$, uses the properties of a slow magnetic diffusion to reduce the number of time steps. In particular, the trapped flux can be calculated with only two time-steps: during the first step, the applied magnetic flux density is increased with a constant sweep rate to a maximum value, it then decreases to zero with the same sweep rate during the second step.
\item[(3)]{\bf 3D finite element model}\\
The last model is based on the same equations as those of model (2), but solves them for a three-dimensional sample with a finite height. Details of the model and its numerical computation are found in Ref.~\cite{Lousberg2}; in particular, the two time-steps method is also used.
\end{enumerate}

Models (1) and (2) are 2D models that ignore the finite height of the samples. Such models are efficient in reproducing the trapped magnetic flux in the median plane of a sample, provided its height is larger than the characteristic length of its cross-section~\cite{Sam,Brandt}. This condition is not quite fulfilled for the samples under study. Moreover, the 2D models only simulate the trapped flux in the median plane, while we only have access to the value at the surface with experiments. However, these models are believed to provide a first approximation of the relative magnetization drop in drilled samples in a much faster way than the 3D model does, without requiring excessive calculation loads. These are the reasons why the 2D models are considered in this paper.

\section{Measurement of the trapped magnetic flux density}
\label{s:measure}

\begin{figure}[b!]
\center
\includegraphics[width=10cm]{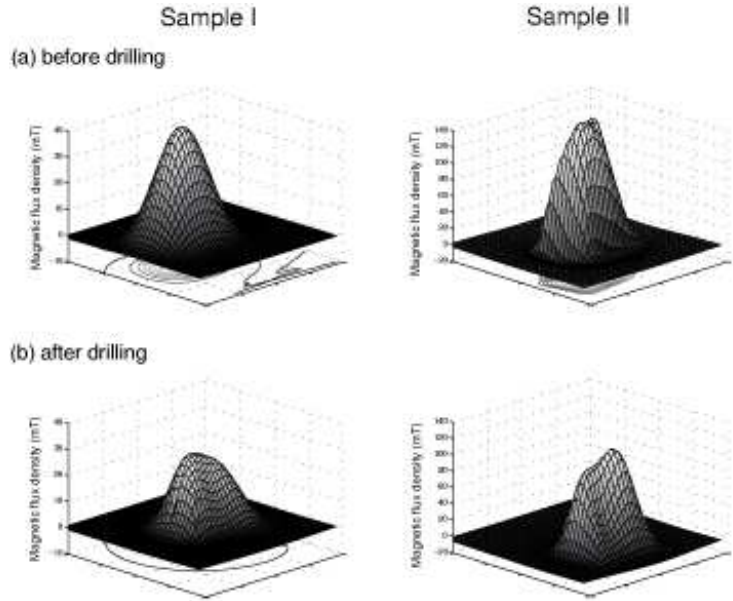}\caption{Distribution of the trapped magnetic flux density at $\approx0.5~\mathrm{mm}$ above the sample surface of samples I and II, before and after drilling.}\label{FigureMapping3D}
\end{figure}

The trapped magnetic flux density above the sample surface was measured in a Hall probe mapping experiment, by moving a miniature probe fixed to a motor-driven $xy$ micro-positioning stage over the sample surface~\cite{FirstDrilled}. The Hall probe is sensitive to the component of the local field which is perpendicular to the surface. Its active area is $0.05\times0.05~\mathrm{mm}^2$ and its distance from the sample surface is $0.5~\mathrm{mm}$. The Hall probe was moved across the top surface with a step size of $0.5~\mathrm{mm}$ in $x$ and $y$ directions. In addition, we also measured the maximum trapped magnetic flux density directly on the surface. We brought the Hall probe in close contact with the sample and moved it slightly around the sample centre until the maximum reading was found.

The samples were magnetized with a field-cooling process during which the samples, initially at room temperature, were immersed in liquid nitrogen in a uniform magnetic flux density of $300~\mathrm{mT}$ created by a large copper coil. The field-cooling process lasted $5~\mathrm{min}$, and the characterization of the trapped field started $15~\mathrm{min}$ after the applied magnetic field had been switched off, so that magnetic relaxation effects were negligible.

The distribution of the trapped magnetic flux density at $0.5~\mathrm{mm}$ above the top surface of sample I --- left panels --- and sample II --- right panels --- are shown in Figure~\ref{FigureMapping3D}, respectively before drilling, (a), and after drilling, (b). In sample I, the trapped magnetic flux density has a single maximum at $B_{\mathrm{trapped}}^{\mathrm{max}}\approx 40~\mathrm{mT}$ before drilling; after drilling, the maximum field drops to $B_{\mathrm{trapped}}^{\mathrm{max}}\approx 25~\mathrm{mT}$. In sample II a larger maximum trapped flux density is found before drilling, with $B_{\mathrm{trapped}}^{\mathrm{max}}\approx 130~\mathrm{mT}$; it reduces to $B_{\mathrm{trapped}}^{\mathrm{max}}\approx 100~\mathrm{mT}$ after drilling. The relative drop of the maximum trapped flux density is found to be larger in sample I. It can also be observed that, in addition to reducing the trapped flux, holes also modify its distribution with the largest effects near the maximum.

\section{Discussion}
\label{s:discussion}

\begin{figure}[b!]
\center
\includegraphics[width=11cm]{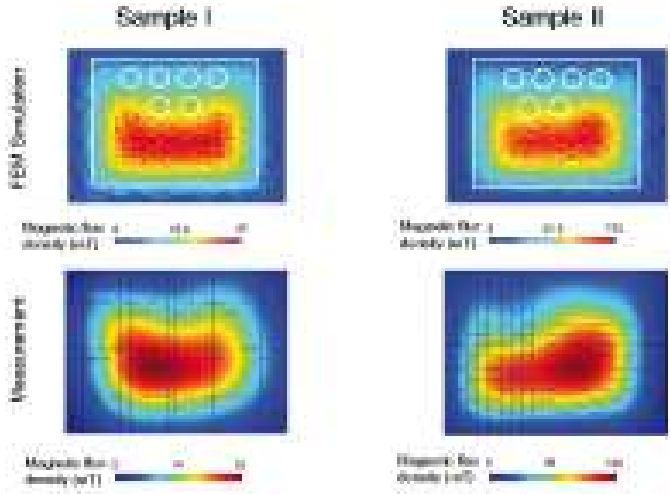}\caption{Two-dimensional trapped magnetic flux distribution at $\approx0.5~\mathrm{mm}$ above the sample surface of the drilled samples I and II. The top panels correspond to the simulation results with the 3D-FEM models and the bottom panels to the measurements. }\label{FigureMapping2D}
\end{figure}

The measured drops in trapped flux of the drilled samples are compared to the simulation results. The same set of parameters are used for drilled and plain samples. They are fixed as follows. First, the critical current density is chosen such that model (3) reproduces well the value of the maximum trapped flux density that is measured on the top surface of each plain sample. This gives $J_c=4.1~10^{7}~\mathrm{A/m}^2$ for sample I and $J_c=8.8~10^{7}~\mathrm{A/m}^2$ for sample II. Second, for the $E-J$ power law in models (2) and (3), the critical exponent $n$ is taken as $n\approx25$, a typical value for YBCO bulk samples~\cite{nexpo}. 

Unlike the experiments where the samples are magnetized with a field-cooling process, the simulations reproduce a zero field-cooling process, and hence require a maximum applied field larger than $300~\mathrm{mT}$ as used in experiments. In order to fully magnetize the samples, even those with an infinite height, the maximum value of the applied magnetic flux density is chosen at $1~\mathrm{T}$ and the sweep rate is equal to $1~\mathrm{mT/s}$. 
Let us first compare the distributions of the trapped flux above the surface of the drilled samples with that calculated with the 3D FEM model. Figure~\ref{FigureMapping2D} shows the measured trapped flux distribution at $0.5~\mathrm{mm}$ above the surface in the bottom panels and the numerical predictions in the top panels, for each sample. Note that in the experimental results, the holes cannot be resolved because of the distance between the Hall probe and the surface of the samples. In sample I, the simulated distribution of the trapped flux is symmetric, with a maximum trapped flux density occuring at the center of the cross section. This distribution deviates from the experimental mapping (bottom left panel), where the maximum trapped field is found to be slightly shifted to the left of the centre.  The agreement between the calculated and the measured distributions is much better for sample II, where the maximum of the trapped flux is found to be shifted to the right of the centre in both the experimental mapping and the simulated distribution.

\begin{table}[t!]
\caption{\label{TableSimul1}Comparison of the maximum trapped magnetic flux density in samples I and II}
\begin{center}
\renewcommand{\arraystretch}{1.4}
\begin{tabular}{rccc|cc}
\br
{\bf Sample I}& \multicolumn{3}{c}{\textsc{Median plane}} & \multicolumn{2}{c}{\textsc{Top surface}}\\
&\textsc{Bean model}&\textsc{FEM - 2D}&\textsc{ FEM - 3D} & \textsc{ FEM - 3D} & \textsc{Measurement}\\
\hline
Before drilling& $169~\mathrm{mT}$ & $137~\mathrm{mT}$ & $95~\mathrm{mT}$ & $61~\mathrm{mT}$ & $60~\mathrm{mT}$ \\
After drilling& $126~\mathrm{mT}$ & $104~\mathrm{mT}$ & $75~\mathrm{mT}$ & $46~\mathrm{mT}$ & $33.7~\mathrm{mT}$\\
\hline Relative drop & $25\%$&$24\%$ & $21\%$ & $25\%$&$44\%$ \\
\br
\end{tabular}
\begin{tabular}{rccc|cc}
\br
\vspace{0.1cm}{\bf Sample II}&\multicolumn{3}{c}{\textsc{Median plane}} & \multicolumn{2}{c}{\textsc{Top surface}}\\
&\textsc{Bean model}&\textsc{FEM - 2D}&\textsc{ FEM - 3D}& \textsc{ FEM - 3D} & \textsc{Measurement}\\
\hline
Before drilling& $358~\mathrm{mT}$ & $310~\mathrm{mT}$ & $244~\mathrm{mT}$& $154~\mathrm{mT}$ & $155~\mathrm{mT}$  \\
After drilling& $291~\mathrm{mT}$ & $253~\mathrm{mT}$ & $207~\mathrm{mT}$& $130~\mathrm{mT}$ & $120.7~\mathrm{mT}$\\
\hline Relative  drop& $19\%$&$18\%$ & $15\%$ & $16\%$&$22\%$\\
\br
\end{tabular}
\end{center}
\end{table}

The measurement of the trapped flux distribution at a finite distance above the sample surface can be affected by a systematic error in the distance between the Hall probe and the sample surface that cannot be reproduced accurately in simulations. It is thus interesting to compare the maximum trapped flux densities when the probe is in contact with the sample. Table~\ref{TableSimul1} reports the simulated and the measured maximum trapped flux density in sample I and II, before and after  drilling. 

The maximum trapped flux density is simulated in the median plane of the samples with models (1), (2) and (3) and  on the surface with model (3). It is found that the maximum trapped flux density is in each case the largest with the Bean model, the flux creep and geometrical effects bringing additional drops of trapped flux. Further, although the drops in magnetic flux density are of different magnitudes in each model, the relative drops are found to be nearly equal,  about $23~\%$ for sample I and $17\%$ for sample II. These results indicate that the centered rectangular lattice is the configuration with the smallest drop in the trapped flux, as was already found in samples with different cross sections in Ref.~\cite{Lousberg}. 

Experimentally, the measured maximum magnetic flux density is found to drop by $44\%$ in sample I, and by $22\%$ in sample II. On the one hand, the smallest relative drop that is observed in sample II is well reproduced by simulations, which confirm the previous theoretical results in Refs.~\cite{Lousberg,Lousberg2}. It is interesting to note that the simple 2D models, while not absolutely suitable for samples under consideration in this study, already yield a reasonable estimate of the relative drop that is measured at the surface of sample II. For sample I, on the other hand, the numerical simulations fail in describing the large measured drop.

\begin{figure}[t!]
\center
\includegraphics[width=12cm]{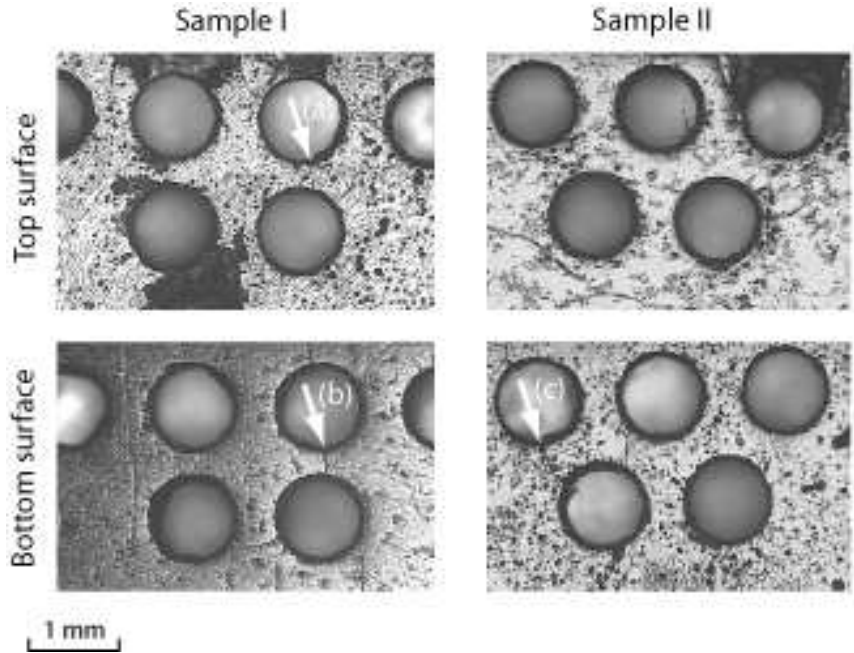}\caption{Micrographs of the bottom and top surfaces of sample I and II after the drilling of the holes.}\label{FigureCracks}
\end{figure}

Given the discrepancy between theory and experiment for sample I, the large observed field drop cannot be only due to a reorganisation of the current lines after drilling. Presumably, drilling also affected the microstructure of the sample. Figure~\ref{FigureCracks} shows optical microscope pictures of top and bottom surfaces of sample I and II after drilling. The arrows (a) and (b) show a crack between two neighboring holes in sample I. This crack is found on both surfaces and has probably be produced during drilling. Its position is consistent with the  shift to the left observed in the maximum of the field distribution in Figure~\ref{FigureMapping2D}. By contrast, sample II presents a crack on one of its sides (arrow (c)) but not on the other, so that drilling appears to have preserved most the microstructure of this sample.

\section{Conclusion}
\label{s:concl}

We have shown with experiments and modelling that the arrangement of the holes in a drilled sample influences the trapped magnetic flux. We considered two samples with the same dimensions and made six holes in each, according to two different arrangements.  In sample I, the holes are placed on a rectangular lattice, and in sample II, they form a centered rectangular lattice.

The measurements of the trapped flux density above the sample surfaces have shown that the drilled samples trap a smaller magnetic flux than a sample with no hole. The sample II with the centered rectangular lattice exhibits the smallest relative drop of the maximum trapped magnetic flux. 

The measurements have been compared to the results obtained with numerical simulations. We have considered three models: (1), a 2D Bean model neglecting the flux creep and assuming that the samples have an infinite height, (2), a 2D FEM model with an $E-J$ power law, and (3), a 3D FEM model taking into account both the flux creep and the finite size of the samples. A good agreement between the simulations with each of the three models and the measurements has been found for the relative drop of trapped flux in sample II. Sample I exhibits a larger drop than what is predicted by the simulations, most probably because drilling produced cracks over the full sample height.

\section{Acknowledgments}
Authors would like to thank the University of Liege (ULg) and the Belgian {\it Fonds de la Recherche Scientifique} (FRS-FNRS) for cryofluid and equipment grants. They are also grateful to Jacques Noudem (CRISMAT - Caen) for the synthesis of the samples.

\end{document}